\documentclass[11pt,a4paper]{article}

\pdfoutput=1
\usepackage{jheppub}

\usepackage{epsfig}
\usepackage{hyperref}
\usepackage{amssymb}
\usepackage{amsbsy}
\usepackage{amsmath}
\usepackage{url}

\newcommand{\rf}[1]{(\ref{#1})}
\newcommand{\beq}{\begin{equation}}
\newcommand{\eeq}{\end{equation}}
\newcommand{\be}{\begin{equation}}
\newcommand{\ee}{\end{equation}}
\newcommand{\bea}{\begin{eqnarray}}
\newcommand{\eea}{\end{eqnarray}}
\newcommand{\eq}[1]{Eq.~(\ref{#1})}
\newcommand{\non}{\nonumber \\*}
\newcommand{\ie}{{i.e.}\ }
\newcommand{\bv}{\begin{verbatim}}
\newcommand{\ev}{\end{verbatim}}

\newcommand{\vp}{\varphi}

\newcommand{\e}{\,\mbox{e}}
\renewcommand{\d}{{\rm d}}
\renewcommand{\i}{{\rm i}}
\newcommand{\blambda}{\bar\lambda}
\newcommand{\brho}{\bar\rho}

\newcommand{\half}{{\textstyle \frac 12}}

\newcommand{\bz}{{\bar z}}
\newcommand{\p}{\partial}
\newcommand{\bp}{\bar\partial}
\newcommand{\q}{\mbox{}}
\newcommand{\qq}{q}
\newcommand{\hg}{{\hat g}}

\newcommand{\eps}{\varepsilon}
\newcommand{\om}{\omega}
\newcommand{\del}{\delta}

\newcommand{\LA}{\left\langle}
\newcommand{\RA}{\right\rangle}

\def\fun#1#2{\lower3.6pt\vbox{\baselineskip0pt\lineskip.9pt
\ialign{$\mathsurround=0pt#1\hfil##\hfil$\crcr#2\crcr\sim\crcr}}}

\begin{document}

\preprint{}

\title{Singular products and universality in higher-derivative conformal theory}

\author
{Yuri Makeenko}
\vspace*{2mm}
\affiliation{NRC ``Kurchatov Institute''\/-- ITEP,  Moscow\\
}
\emailAdd{makeenko@itep.ru} 

\abstract
{I investigate universality of the two-dimensional  higher-derivative conformal theory
using the method of singular products. 
The previous results for the central charge at one loop are confirmed for the quartic
and six-derivative actions.}


\maketitle


 \section{Introduction}
 
 Corformal invariance of Polyakov's string formulation is widely used since 1980's
 (see~\cite{Pol87}). The simplest observable for a closed string without external sources
 is then the string susceptibility index $\gamma_{\rm str}$ 
 (also known as the gravity anomalous dimension) which is defined 
 through the number of surfaces of large area $A$ 
 as
\be
\LA \delta \left(\int \sqrt{g} -A \right) \RA \propto A^{\gamma_{\rm str}-3} \e^{C A}.
\ee
Here the constant $C$ in the exponent is not universal, \ie depends on the regularization
procedure applied to the string, but the pre-exponential is perfectly universal and
describes a gravitational dressing of the unit operator.

 The celebrated calculation of 
$\gamma_{\rm str}$ was performed
by Knizhnik-Polyakov-Zamolodchikov \cite{KPZ} (KPZ) by fixing the light-cone gauge
and by 
David~\cite{Dav88}, Distler-Kawai~\cite{DK89} (DDK) using the conformal gauge
with the result
\be
\gamma_{\rm str}=(1-h)\left[\frac{d-25-\sqrt{(25-d)(1-d)}}{12}\right]+2
\label{ggg}
\ee
for a surface of genus $h$ embedded in $d$ Euclidean dimensions.
Equation~\rf{ggg} extends  the one-loop result~\cite{Zam82,CKT86,KK86} to all
orders  in $1/d$ (as  $d\to-\infty$) and describes critical indices of the vast amount of
models in Statistical Mechanics where the central charge $c=d<1$. 
However, as is seen from \eq{ggg},  $\gamma_{\rm str}$ is not real
for $1<d<25$ as it should say for the QCD string in $d=4$. 
This was referred to as the $d=1$ barrier for the string existence.

The derivation of \eq{ggg} is based on the Liouville action which emerges
from the Polyakov string after doing the path integral over all fields but
the independent world-sheet metric tensor $g_{ab}$. The integration over the 
target-space coordinates $X^\mu$ is 
commonly done by the DeWitt-Seeley expansion of the heat kernel~\cite{deWitt,Gil75}
\be
\LA \omega \Big| \e^{\tau \Delta }
\Big| \omega \RA = \frac 1{4\pi\tau}
 +\frac1{24\pi} {R(\omega)}+\frac{\tau}{120\pi}
\left[ \Delta R(\omega)+\frac 12 R^2(\omega)
\right]+{\cal O}(\tau^2),
\label{Seeley}
\ee
where $\Delta$ denotes the two-dimensional Laplacian and $R$ is the scalar curvature
for the metric tensor $g_{ab}$.
The conformal gauge is fixed by choosing 
\be
g_{ab}=\hg _{ab}\e^\vp 
\label{confog}
\ee
with $\hg _{ab}$ being the background (also termed fiducial) metric tensor and
$\vp$ being a dynamical variable often termed the Liouville field.
Also the ghosts have to be added when fixing the gauge \rf{confog}.

Equation~\rf{Seeley} results after the path-integrating over $X^\mu$  in the
piece of the {\em emergent} action for $\vp$ which is proportional to $d$.
Truncating the DeWitt-Seeley expansion at the first four terms shown in \eq{Seeley} and
adding the contribution from the ghosts, we arrive at the following emergent action for
the Polyakov string:
\be 
{\cal S}^{\rm Pol}=\frac 1{16 \pi b_0^2} \int \sqrt{\hg} \left[\hg^{ab} \p_a \vp \p_b \vp  +2\mu_0^2\e^\vp
+\eps  \e^{-\vp}(\hat \Delta \vp)^2 \right] 
, \qquad
b_0^2=\frac6{26-d}.
\label{invaR}
\ee
Here the first two terms forming the Liouville action are familiar from the original work by
Polyakov~\cite{Pol81}. The remaining third term 
is familiar from the studied~\cite{KN93,Ich95,KSW} of $R^2$ gravity in two dimensions.
The constant $\eps\propto\tau/\sqrt{\hg}$  originates from two contributions. 
The first comes from the path integration
over $X^\mu$ and is proportional to $d$. It is easily calculable from the last two terms shown in the 
DeWitt-Seeley expansion~\rf{Seeley}. The second comes from the path integral over the ghosts.
I shall give more details on this issue in the next section.

An analogue of the emergent action~\rf{invaR} can be derived also for the Nambu-Goto string.
Doing again the path integral over all fields but $g_{ab}$, we arrive 
in the conformal gauge~\rf{confog} at the emergent action
of the type~\rf{invaR} but with the additional term
\be 
{\cal S}=\frac 1{16 \pi b_0^2} \int \sqrt{\hg} \left\{\hg^{ab} \p_a \vp \p_b \vp  +2\mu_0^2\e^\vp
+\eps  \e^{-\vp}\left[(\hat \Delta \vp)^2 - G \hg^{ab}\,\partial_a \vp\partial_b \vp  
\hat \Delta\vp\right] \right\} 
 .
\label{inva}
\ee
It is the most general diffeomorphism-invariant action with four derivatives.
All other terms with four derivatives can be reduced to these two (modulo boundary terms
in the case of an open string).  The term with $G$ does not appear for the Polyakov string.
Its occurrence is specific~\cite{Mak21} to the Nambu-Goto string.
How to compute the value of $G$ for the Nambu-Goto string will be outlined
in Sect.~2 and Appendix~A. I shall not concentrate however at that particular value 
of $G$ and rather consider the  action~\rf{inva} as such.

Strictly speaking \eq{inva}, as derived by the path integration over all the fields but $g_{ab}$,
holds only for flat backgrounds when $\hat R$ -- the scalar curvature for the metric
tensor $\hat g_{ab}$ -- vanishes. As is well-known, under the decomposition~\rf{confog}
one has
\be
\sqrt g R= \sqrt{\hat g} \left( \q\hat R - \hat \Delta \vp \right)
\label{Rshift}
\ee
with $\hat \Delta $ being the Laplacian for the metric tensor $\hg_{ab}$,
which results in the appearance of the addition
$\propto\!\vp \hat R$ to the Liouville action and similar terms for the four-derivative
 action~\rf{inva}, causing a non-minimal interaction of $\vp$ with background gravity.
These additional terms are crucial for the Weyl invariance of the four-derivative action~\rf{inva}
as well as for the derivation of the ``improved'' energy-momentum tensor in flat space.
I shall return to this issue in Subsect.~\ref{ss:3.1}.
The action~\rf{inva} is thus conformal invariant in flat space in spite of the presence of
the dimensionful parameters $\mu_0^2$ and $\eps$, 
so the methods of conformal field theory can be applied.

The parameter $\tau$ of the DeWitt-Seeley
expansion~\rf{Seeley} is actually Schwinger's proper-time regularization
of ultraviolet divergences in the path
integral. Thus $\eps$ being proportional to the target-space cutoff $\tau$
is negligibly small. For this reason the four-derivative terms in the action~\rf{inva} 
are classically suppressed for smooth metrics as $\eps R$. However, the role 
 of the parameter $\eps$  in the quantum case is twofold. Firstly, 
the quartic derivative regularizes divergences with $\eps$
playing the role of an {ultraviolet worldsheet cutoff}. Secondly, $\eps$ is simultaneously 
a coupling constant of the
self-interaction of $\vp$ so  uncertainties like $\eps \times \eps^{-1}$ appear in the
perturbation theory. In other words,
typical metrics essential in the
path integral over $g_{ab}$ are not smooth and have $R\sim \eps^{-1}$. 
These uncertainties look like anomalies in quantum field theory and may affect the large-distance
behavior of strings as argued in~\cite{Mak21}.
 
In particular, the string susceptibility index 
computed for the four-derivative
action \rf{inva} with the one-loop accuracy equals~\cite{Mak22,Mak22c} 
\be
\gamma_{\rm str}=
(h-1)\left( \frac {1}{b_0^2}  -\frac76-G +{\cal O}(b_0^2)\right) +2,\qquad 
b_0^2=\frac6{26-d}
\label{gstrfin}
\ee
for closed surfaces of genus $h$,
showing for $G\neq0$ a deviation from the one-loop result~\cite{Zam82,CKT86,KK86}
for the Polyakov string for which $G=0$.

Equation~\rf{gstrfin} was derived from the action~\rf{inva} in three different ways.
Firstly, the  technique \cite{KPZ,Dav88,DK89} of conformal field theory 
developed for the Liouville action was
applied  to compute the central charge of $\vp$
at one loop and, secondly, the results where confirmed~\cite{Mak22} 
by a direct computation of
 the one-loop diagrams of quantum field theory, which describe the renormalization of the
 propagator and the energy-momentum tensor.
The third method has been recently proposed~\cite{Mak22c} as a pragmatic mixture of the two, 
accounting for the quantum equation
of motion. It has also reproduced~\eq{gstrfin} via emerging singular products.

We see from \eq{gstrfin} that the string susceptibility for the four-derivative action~\rf{invaR}
coincides at one loop with the one for the Liouville action. A natural question is as to whether this
holds to all loops? Another natural question is what about a more general action
of the form 
\be
{S}^{\rm gen} [\vp]= -\frac 1{16\pi b_0^2}\int  \sqrt{\hg} 
 \vp \hat\Delta F( -\eps\e^{-\vp}\hat\Delta) \vp,\qquad F(0)=1
\label{Sgen}
\ee
which differs from the action~\rf{invaR} by the terms of order $\eps^2$ and higher.
The action~\rf{Sgen} has been discussed recently in~\cite{Mak21,ST22}. 
It arises by covariantizing a free higher-derivative action which is
quadratic in $\vp$ and therefore modifies the propagator.  
The function $F$ depends on the applied regularization and has been computed~\cite{Mak23b} to all orders in $\eps$. 
The action~\rf{Sgen} is a part of the most general
higher-derivative action generated by yet higher-order terms of the DeWitt-Seeley  
expansion of the heat kernel after the
path-integration over $X^\mu$ and the ghosts in the Polyakov string formulation.
The difference between~\rf{Sgen} and the most general action
shows up already at the order $\eps^2$, 
where the term $\eps^2 R\Delta R$ is captured by~\rf{Sgen} but the term $\eps^2 R^3$ is not.

This Paper is addressed to the question of the universality, \ie an independence of 
the central charge and
$\gamma_{str}$ on the precise form of the higher-derivative emergent action.
In analyzing this I found it most useful to apply the method based on the singular products 
which has been recently developed in~\cite{Mak22c}.
After a brief reviewing of the subject in Sect.~2, I concentrate 
on the universality for higher-derivative actions. It is demonstrated 
in Sects.~\ref{s:3} and \ref{s:4} by showing that the contribution to the central
charge of $\vp$ from the four-derivative  
$R^2$  and six-derivative $R\Delta R$ or $R^3$ terms 
vanishes, so only the usual one coming from the quadratic part remains.
How the universality works for the action~\rf{Sgen} is illustrated by an example
in Sect.~\ref{s:5}.
I also apply the method of singular products to confirm \eq{gstrfin} by
computing in Subsect.~\ref{ss:3.3} the central charge of $\vp$ at one loop via 
the variation of the energy-momentum
tensor under an infinitesimal conformal transformation.
Appendix~\ref{AppA} is devoted to the computation of $G$ in \eq{inva} as emerging
from the Nambu-Goto string.
Numerous formulas for the singular products are derived in Appendix~\ref{AppB}.

\section{Preliminaries and the setup}

Let us begin with reminding the relation between the Polyakov and Nambu-Goto 
string formulations.
The action of the Polyakov string
\be
S=\frac{K_0}{2}\int \sqrt{g} g^{ab} \p_a X \cdot \p_b X,
\label{SPol}
\ee
where $K_0=1/2\pi\alpha'_0$ stands for the bare string tension,
is quadratic in $X^\mu$ that makes it easy to integrate it out in the path integral.
The world-sheet metric tensor $g_{ab}$ is an independent field in the path integral.
The Nambu-Goto action of the bosonic string is the area
of the string worldsheet. It is highly nonlinear in $X^\mu$ but
can be made quadratic introducing the (imaginary) Lagrange multiplier 
$\lambda^{ab}$ and an independent metric tensor $g_{ab}$ as
\be
S_{\rm NG}=K_0\int \sqrt{\det {(\p_a X\cdot \p_b X)}}=
K_0\int \Big [\sqrt{g} +\frac 12\lambda^{ab} (\p_a X\cdot \p_b X-g_{ab}) \Big].
\label{SNG}
\ee
In both cases it is convenient to diagonalize $g_{ab}$,
choosing the conformal gauge~\rf{confog}.
This procedure adds ghosts which are the same for both string formu\-lations.

In the classical limit when $\alpha'_0\to 0$ or $K_0\to\infty$ we have
\be
g_{ab}^{({\rm cl})} =\p_a X\cdot \p_b X,
\label{gcl}
\ee
\ie it coincides with the induced metrics. Analogously $\lambda^{ab}=\sqrt{g}g^{ab}$
for the classical ground state.
The substitution of \rf{gcl} into the
action~\rf{SPol} then reproduces the Nambu-Goto action~\rf{SNG}. 
It was also demonstrated~\cite{FTs82} that both string formulations
give the same results at one loop providing the zeta-function regularization is used.
The general argument in favor of the equivalence of the two string formulations is
 based~\cite{Pol87} on
the fact that $\lambda^{ab}$ is localized at the distances of the order of the ultraviolet (UV)
cutoff and does not propagate to macroscopic distances. I shall return soon to this argument.

A separate remark is required about the stability of the classical ground state~\rf{gcl}.
As shown in~\cite{AM16}, it is stable only for $d\leq 2$. For $d>2$ the mean-field
ground state is stable instead for which $g_{ab}= \brho g_{ab}^{({\rm cl})}$ and
$\lambda^{ab}=\blambda \sqrt{g}g^{ab}$ with certain $d$-dependent values of $\brho>1$
and $\blambda<1$ well-defined for $d>2$. The continuum limit is reached in
the scaling regime when the
UV cutoff is going away. Then
$\brho\to\infty$ forcing the stringy continuum limit to be Lilliputian~\cite{AM16b}.
I shall not touch upon that issue  in this Paper which deals with the one-loop approximation
justified by $d\to -\infty$ when the ground state is classical. 

As is already mentioned, the main reason to introduce the Lagrange multiplier for the Nambu-Goto string
was to path integrate over $X^\mu$ (as well as over the ghosts) to obtain the
emergent action for the fields $\vp$  and $\lambda^{ab}$. 
The arising determinants diverge and
have to be regularized. What regularization to use for this purpose is a matter of
personal taste.
I prefer to use the covariant Pauli-Villars regularization which was first introduced
for the ghosts~\cite{Dia89}. To regularize $X^\mu$ we then introduce the massive regulators 
$Y^\mu$ which obey wrong statistics and add to \rf{SNG} the regulator action
\be
{S}_{\rm reg}=\frac {K_0}2 \int \left (\lambda^{ab} \p_a Y \cdot\p_b Y
 +M^2\sqrt{\hg} \e^\vp\, Y^2 \right).
\label{Sreg}
\ee
Now every loop of the regulator field $Y^\mu$ brings the minus sign to compensate divergences
coming from $X^\mu$.

 Actually, we have to have~\cite{AM17c} two such regulators 
of mass squared $M^2$ with wrong statistics which can be viewed as anticommuting Grassmann
variables $Y^\mu$ and $\bar Y^\mu$ 
and one regulator $Z^\mu$ of mass squared $2M^2$
with normal statistics to regularize all the divergences including the ones in tadpole
diagrams. But for the purposes of computing the finite parts (like anomalies) only one
regulator will be enough because the contributions of the two others are canceled being
independent of the masses.

For the covariant Pauli-Villars regularization we can apply 
the standard methods of quantum field theory.
It is seen from Eqs.~\rf{SNG} and \rf{Sreg} how the vertices of the
interaction of the fields $\vp$ and $\lambda^{ab}$ with $X^\mu$ and the regulators arise.%
\footnote{For the Polyakov string we substitute $\lambda^{ab}=\sqrt{\hg} \hg^{ab}$ so
only the interaction between $\vp$ and the regulators remains.}
Feynman's diagrammatic technique can be
used for the calulation of the emergent action.
Also Noether's theorems apply to the system of
$X^\mu$ plus the regulators. In particular, the total energy-momentum tensor can
be derived, which is conserved and traceless thanks to the classical equations of motion.
The covariant Pauli-Villars regulators thus preserve conformal symmetry in spite of they
are massive. For this reason the emergent action will be conformal invariant which
is crucial for what follows. The coeffients of the Taylor series of the function $F$ in \eq{Sgen}
and other terms in the emergent action depend
on the regularization applied. For the proper-time and
Pauli-Villars regularizations they are related by simple formulas as shown in~\cite{Mak23b}
(Appendix~A).

\subsection{The emergent action}

The fields $X^\mu$ enter the action~\rf{SNG} quadratically (the same for the ghost fields
in the ghost action) so they can be integrated out. In~\cite{Mak21} the action, governing
fluctuations of $\vp$ and $\lambda^{ab}$, that emerges from the Nambu-Goto string
was analyzed by the use of 
the covariant Pauli-Villars regularization.  
I briefly reiew in this subsection the result and repeat the derivation  by using
the proper-time regularization in Appendix~\ref{AppA}. 

\begin{figure}
\centerline
{\includegraphics[width=9cm]{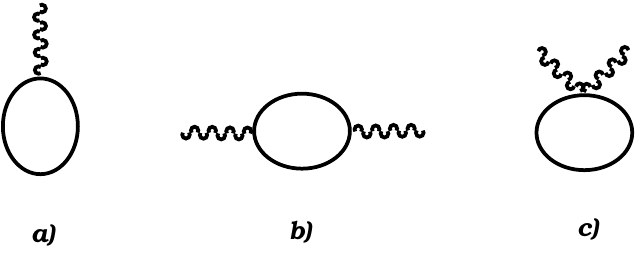}}
\caption{Diagrams contributing to the emergent action to quadratic order in $\vp$ or
$\lambda^{ab}$ represented by the wavy lines. The solid line represents the loop of either
$X^\mu$ or the regulators.}
\label{FIGtada}
\end{figure}

The diagrams which contribute after the path integration over $X^\mu$ and its regulators
to the emergent action to quadratic order in $\vp$ or
$\lambda^{ab}$ are depicted in Fig.~\ref{FIGtada}. The 
 wavy lines represent either $\vp$ or $\lambda^{ab}$ while the solid line represents 
 the loop of either $X^\mu$ or its regulators. 
Covariantizing, we arrive at the following emergent action of the Nambu-Goto string:
\bea
\lefteqn{{\cal S}_X[\vp, \lambda^{ab}] }\non
&&= \frac d2 \int \left[ -\frac {\sqrt{\hg}\e^{\vp}\Lambda^2}{ \sqrt{\det{\lambda^{ab}}}}
+\frac1{48\pi}\left( \sqrt{\hg}\vp \hat\Delta\vp +
\lambda^{ab} \hg_{ab}\hat \Delta \vp+2
\lambda^{ab}\nabla_a\p_b  \vp  \right)
+\frac{\sqrt{\hg}\e^{-\vp}}{160\pi M^2}  (\hat\Delta \vp)^2 \right] \non
&& \hspace*{1cm}+ {\cal O}(M^{-4}),\qquad \quad
\qquad\Lambda^2=\frac{M^2}{2\pi}\log 2
\label{SXb}
\eea
with $\nabla_a$ being the covariant derivative for $g_{ab}$ given by \eq{confog}.
Expanding $1/ \sqrt{\det{\lambda^{ab}}}$ in the fluctuating part 
$\delta \lambda^{ab}=\lambda^{ab}-\sqrt{\hg} \hg^{ab}$ [cf.~\eq{detla}],
we see from the action~\rf{SXb}  that $\delta \lambda$'s have
the mass squared $\propto \! \tau^{-1}$ and are therefore localized at the distances
$\sim \sqrt{\tau}$.

The contribution from the ghosts associated with fixing the conformal gauge
is just the same as for the Polyakov string 
\be
{\cal S}_{\rm gh}[\vp]= \int \sqrt{\hg}\e^{\vp}\left[ \Lambda^2
 -\frac {13}{48\pi}\vp \Delta\vp -\frac{11}{160\pi M^2}  (\Delta \vp)^2 \right]
 + {\cal O}(M^{-4}).
\label{Sgh}
\ee
The fist two terms on the right-hand side are well-known and the third one has been
recently computed~\cite{Mak23b}.

Path-integrating over $\lambda^{ab}$ by the saddle-point method as described in 
Appendix~\ref{AppA}, using the identity~\rf{ide4} and dropping the terms
${\cal O}(M^{-4})$, we reproduce 
the four-derivative action~\rf{inva}.

\subsection{``Improved'' energy-momentum tensor}
 
The  $T_{zz}$ component of
the energy-momentum tensor associated with
the action~\rf{inva} reads~\cite{Mak22}
\bea
-4 b_0^2 T_{zz}&=&(\p \vp)^2 -2\eps \p \vp \p \Delta \vp -2\q\p^2 (\vp - \eps\Delta \vp)
-G \eps (\p \vp)^2 \Delta \vp+ 4G\eps\p\vp \p(\e^{-\vp}\p \vp \bp \vp)  \non
 && 
- 4G\q\eps \p^2(\e^{-\vp}\p \vp \bp \vp)
 +  G\q \eps\partial(\p \vp \Delta\vp)+
  G \q\eps \frac 1{\bp}\p ^2 (\bp \vp \Delta\vp) ,
  \label{Tzz}
  \eea
where $\Delta=4\e^{-\vp}\p  \bp$ when using the conformal coordinates $z$ and $\bz$
for a flat metric tensor $\hat g_{ab}$.
We have used here the notation $\partial\equiv \partial/\p z$ and
$\bar \partial\equiv \partial/\p\bz$. 
Notice the nonlocality of the last term in \rf{Tzz} which is inherited from a
nonlocality of the covariant generalization of the action \rf{inva}. 
It is the presence of this nonlocal term which plays a crucial role in the computation of 
the addition $6G$ to the central charge at one loop.

It is important that the energy-momentum tensor~\rf{Tzz} is ``improved'' 
\'a la Callan-Coleman-Jackiw~\cite{CCJ70,DJ95}.
It is conserved and traceless 
owing to the classical equation of motion 
despite  $\eps$ is dimensionful. This is  a consequence of diffeomorphism invariance
of the action \rf{inva} which
thus possesses conformal symmetry at
least at the classical level.

The action \rf{inva} generates the vertex
 \be
 \LA \vp(k)\vp(p)\vp(q) \RA_{\rm truncated} =\frac\eps{8\pi b_0^2}
 \left( k^2 p^2+k^2 q^2 +p^2 q^2 \right) \delta^{(2)}(k+p+q)
 \ee
which depend on  momenta, so some diagrams diverge
inspite of the $k^4$ in the propagator. This produces generically quadratic divergences in
(tadpole) diagrams of perturbation theory.

To regularize the divergences  we
 implement the covariant Pauli-Villars regularization \cite{Dia89},  adding to \rf{inva} 
 the following action for the regulator field $Y$:
\be
{\cal S}^{(Y)} =\frac 1{16\pi b_0^2}\int \sqrt{\hg}\left\{ 
\hg^{ab}\partial_a  Y \partial_b Y +M^2\e^\vp Y^2+
\eps \e^{-\vp}\big[
(\hat \Delta Y)^2 -G\eps \hg^{ab}\partial_a  Y \partial_b Y \hat\Delta \vp\big]  \right\} .
\label{S0reg}
\ee
It has a very large mass $M$ and obeys wrong statistics to produce 
the minus sign for every loop, regularizing devergences coming from the loops of $\vp$.
I use in \eq{S0reg} the same letter $Y$  as for the regulator $Y^\mu$ in \eq{Sreg} but
this should not cause any problems.

Once again, the introduction of one regulator is not enough to regularize all the divergences.
Some logarithmic divergences still remain. The correct
procedure is to introduce two regulators of mass squared $M^2$ with wrong statistics, 
which can be represented via anticommuting Grassmann variables, and one regulator 
of mass squared $2M^2$ with normal statistics. Then all diagrams including quadratically divergent
tadpoles will be regularized. However, for the purposes of computing final parts one regulator $Y$ would be enough because the contributions of the two others are canceled being
independent on the masses.

The contribution of the regulators to the $T_{zz}$ component of the energy-momentum tensor 
for the action~\rf{S0reg} reads
\be
-4 b_0^2 T_{zz}^{(Y)}=\p Y \p Y
 -2\eps \p Y \p \Delta Y 
-G \eps \p Y\p Y \Delta \vp 
 + 4G\eps\p\vp \p(\e^{-\vp} \p Y \bp Y)  -4 G \q\eps \p^2 ( \e^{-\vp} \p Y \bp Y).
\label{Tzzreg}
 \ee
The total one is the sum of \rf{Tzz} and \rf{Tzzreg}. The total energy-momentum tensor 
 is conserved and traceless thanks to the classical equations of motion for $\vp$ and $Y$.
Thus the Pauli-Villars regulators are classically conformal fields in spite of they are massive.
For this reason the effective action which emerges after the path-integrating over the regulators
will be conformal invariant. To the quartic order in the derivatives it will be
again of the type in \eq{inva} but with renormalized parameters.

In  the infrared limit  the effective action, 
governing smooth fluctuations of $\vp$, becomes  the Liouville action 
and the effective energy-momentum tensor is quadratic 
\be
T_{zz}^{({\rm eff})}=\frac{1}{2b^2} \left( \qq\p^2 \vp -\frac 12 (\p\vp)^2 \right).
\label{Teff}
\ee
Here $b^2$ is the
renormalization of $\vp$, \ie the change $b_0^2\to b^2$ in the action \rf{inva} and $q$ enters 
the renormalization of $T_{zz}$. The arguments are similar to  
David-Distler-Kawai (DDK)~\cite{Dav88,DK89}. 
In the usual case of the Liouville action where $\eps=0$ 
in \rf{inva} they obey the DDK equation
\be
\frac{6q^2}{b^2}+1=\frac{6}{b_0^2},
\label{DDK1}
\ee
where the left-hand side is the central charge of $\vp$.
For the Polyakov string \eq{DDK1} represents the vanishing of the total central charge.

An analogous computation of the central charge of $\vp$ for the higher-derivative action~\rf{inva}
at one loop results in~\cite{Mak22,Mak22c} 
\be
c^{(\vp)}= \frac{6q^2}{b^2}+1+6 G \left(1-2\int \d k^2 \frac{\eps}{1+\eps k^2} \right)+{\cal O}(b^2_0).
\label{cphi}
\ee
The logarithmic divergence on the right-hand side cancels with the one in the string susceptibility,
so it is finite and given by \eq{gstrfin}.
Both additional finite and divergent parts come from the nonlocal (last) term in \rf{Tzz}.

The emergence of the logarithmic divergence is due to
subtleties in the realization of conformal symmetry generated by the energy-momentum
tensor \rf{Tzz} which is classically not a primary conformal field.
Under the  infinitesimal  conformal transformation
\be
\delta_\xi \vp=\xi'+\xi \p \vp
\label{ctra}
\ee
it changes as
\bea
\delta_\xi T^{(\vp)}_{zz} &=&\frac 1{2 b_0^2} \xi''' + 2 \xi' T_{zz} +\xi \p T_{zz} +
\frac 1{b_0^2}G\eps \e^{-\vp}\Big\{ \xi'''' \bp \vp +\xi''' \big(\p\bp \vp-3 \p \vp \bp\vp\big) \non &&+
\xi''\Big[2 \bp\vp (\p\vp)^2-\p\vp\p\bp\vp-\bp\vp\p^2 \vp \Big]
-\e^\vp\frac1{\bp} \big[ \xi'' \p(\e^{-\vp} \bp\vp \p\bp\vp) \big]\Big\},~~~~~
\label{delTza}
\eea
while the usual definition of the central charge $c$ relies on the transformation law
\be
\delta_\xi T_{zz} =\frac c{12} \xi''' + 2 \xi' T_{zz} +\xi \p T_{zz} 
\label{delTz}
\ee
prescribed for the conserved tensorial field (a descendant of the unit operator).

The difference between the right-hand sides of Eqs.~\rf{delTza} and \rf{delTz} is
due to the presence in the action~\rf{inva} at $G\neq 0$ of the additional term
 involving
the structure $g^{ab}\p_a\vp \p_b\vp$ which is scalar but not primary and 
transforms under \rf{ctra} as
\be
\delta_\xi \left(\e^{-\vp} \p\vp \bp\vp\right) = \xi \p \left(\e^{-\vp} \p\vp \bp\vp\right) +
\xi'' \e^{-\vp} \bp\vp.
\ee
Additional terms  do not appear for $G=0$ because
the scalar curvature $R=-4 \e^{-\vp}\p\bp \vp$ is a primary scalar.
Those also do not appear for $T_{zz}^{(Y)} $ given by \eq{Tzzreg} which obeys
\eq{delTz} with $c=0$.

Averaging \rf{delTza} over $\vp$, we get at one loop the following $\xi'''$ term:
\be
\LA\delta_\xi T_{zz}(0)\RA = \xi''' (0)\left( \frac 1{2 b_0^2} - G \int \d k^2 \frac {\eps}{1+\eps k^2}
\right),
\label{xi3}
\ee
reproducing the   logarithmic divergence in \eq{cphi}.
The logarithmic divergence would 
appear neither in the central charge nor in the string susceptibility within the operator formalism 
where the operators are normal-ordered.%
\footnote{In the Euclidean path-integral formalism the normal product is defined by subtracting all
lower correlators from an operator. For example
$$
:\vp(\om)^3\!:=\vp(\om)^3-3\vp(\om)^2 \LA \vp(\om) \RA-3\vp(\om) \LA \vp(\om)^2 \RA- \LA \vp(\om)^3 \RA
$$
and we have
$$
\frac{\delta}{\delta\vp(z)}:\vp(\om)^3\!:=3\delta^{(2)}(z-\om):\vp(\om)^2\!:\,,
\qquad :\vp(\om)^2\!:=\vp(\om)^2-2\vp(\om) \LA \vp(\om) \RA- \LA \vp(\om)^2 \RA.
$$
\label{foo:2}}
The conformal transformation for such a case of the most general action $S[\vp]$ which
is not quadratic in the fields and whose energy-momentum tensor does not obey
\eq{delTz} can be generated by
\be
\hat \delta_\xi \equiv \int_{D_1} \left(\q \xi' \frac {\delta}{\delta \vp}
 +\xi \p \vp\frac {\delta }{\delta \vp}\right)\stackrel{{\rm w.s.}}
=\int_{C_1}\frac{\d z}{2\pi \i} \xi (z) T_{zz} (z),
 \label{hatdel}
 \ee
where the domain $D_1$ includes the singularities of $\xi(z)$ leaving outside possible singularities of the function $X(\om_i)$ on which $\hat \delta_\xi $ acts and $C_1$ bounds $D_1$. 
The second equality in \rf{hatdel} is understood in the weak sense, \ie under path integrals.
In proving the equivalence  of the two forms 
we have integrated the total derivative 
\be
\bp T_{zz} =-\pi \q \p \frac {\delta S}{\delta \vp} + \pi \p \vp\frac {\delta S}{\delta \vp}
\ee
and used the (quantum) equation of motion
\be
\frac {\delta S}{\delta \vp} \stackrel{{\rm w.s.}}=\frac {\delta}{\delta \vp}.
\label{26}
\ee
Actually, the form of $\hat \delta_\xi$ in the middle  of \eq{hatdel} is primary.
Its advantage over the standard
one on the right is that it takes into account a tremendous cancellation of the diagrams in the quantum case, while there are subtleties associated with singular products. 

In the quantum case we have also an additional effect of the regulator
\be
\LA\hat \delta_\xi X(\om_i) \RA =\LA \int_{D_1} \d^2 z \left(\q \xi' (z)\frac {\delta}{\delta \vp(z)}
 +\xi(z)\p \vp(z)\frac {\delta }{\delta \vp(z)}+\xi(z)\p Y(z)\frac {\delta }{\delta Y(z)}\right)X(\om_i) \RA.
 \label{XX}
\ee
Averaging over the regulators as already discussed, we arrive at the effective action
and the effective energy-momentum tensor \rf{Teff}, describing the infrared limit. 
Equation~\rf{XX} is then substituted by
\be
\LA\hat \delta_\xi X(\om_i) \RA =\LA \int_{D_1} \d^2 z \left(\qq \xi' (z)\frac {\delta}{\delta \vp(z)}
 +\xi(z)\p \vp(z)\frac {\delta }{\delta \vp(z)}\right)X(\om_i) \RA.
 \label{XXq}
\ee
It was shown in Ref.~\cite{Mak22c} how 
 to reproduce
\be
\hat \delta_\xi \e^{\vp(\om)} \stackrel{{\rm w.s.}}=(q-b^2)\xi'(\om)\e^{\vp(\om)} +\xi(\om)\p\vp(\om) \e^{\vp(\om)}
\label{29}
\ee
 for the quadratic action
by \rf{XXq} via the singular products listed in Appendix~\ref{AppB}.

 \section{Central charge via the singular products\label{s:3}}
 
 \subsection{General action and energy-momentum tensor\label{ss:3.1}}
 
For the $2N$-th order in derivatives term in the action~\rf{Sgen} 
\be
S^{(\vp,2N)}= \frac 1{16\pi b_0^2}  \int \sqrt g \vp (-\Delta)^N \vp
\label{sss}
\ee
the ``improved'' energy-momentum tensor reads%
\footnote{These energy-momentum tensors are ``improved''~\cite{CCJ70,DJ95} and 
therefore traceless. They differ for this reason from the ones
in Ref.~\cite{KM16}. Our ``improvement'' procedure also differs from the one~\cite{EMT}
in the free case.} 
\be
T_{zz}^{(\vp,2N)}=\frac1{4b_0^2} (-1)^N
\Big[\sum_{k=0}^{N-1}  (\p \Delta^k \vp)(\p\Delta^{N-k-1} \vp)
-2 \p^2 \Delta ^{N-1}\vp \Big].
\ee
For $N=1,2,3$ we write explicitly
\begin{subequations}
\bea
T_{zz}^{(\vp,2)}&=&-\frac1{4b_0^2}  \left[  (\p\vp)^2-2 \p^2\vp\right],
\label{T2}\\
T_{zz}^{(\vp,4)}&=&\frac1{2b_0^2} \left[  (\p\vp)(\p\Delta\vp)-\p^2\Delta\vp\right],
\label{T4}\\
T_{zz}^{(\vp,6)}&=&-\frac1{4b_0^2} \Big[2(\p  \vp)(\p \Delta^2 \vp)+(\p \Delta\vp)^2
-2 \p^2 \Delta ^{2}\vp\Big] .
\label{T6}
\eea
\label{Ts}
\end{subequations}
\!They obey the conservation law
\be
\bp T_{zz}^{(\vp,2N)}=0
\ee
thanks to the classical equations of motion
\begin{subequations}
\bea
-\Delta \vp&=&0, \\
\Delta^2\vp-\frac 12 (\Delta\vp)^2&=&0,\\
-\Delta^3 \vp +(\Delta \vp )\Delta^2 \vp&=&0
\eea
\end{subequations}
demonstrating tracelessness of $T_{ab}^{(\vp,2N)}$.

For the regulator $Y$ we have analogously
\be
S^{(Y,2N)}= \frac 1{16\pi b_0^2} \int \sqrt g Y (-\Delta)^N Y
\ee
and
\be
T_{zz}^{(Y,2N)}=\frac1{4b_0^2} (-1)^N\sum_{k=0}^{N-1} \left[ (\p \Delta^k Y)(\p\Delta^{N-k-1} Y)
\right] .
\ee
The sum $T_{zz}^{(2N)}+T_{zz}^{(Y,2N)}$ is traceless thanks to 
the classical equation of motion for $\vp$.

This is a general property because in the conformal gauge 
\rf{confog}
we have
\be
T^a_a\equiv\hat g^{ab} \frac{\delta {\cal S}[g]}{\delta \hat g^{ab} }= -
\frac{\delta {\cal S}[g]}{\delta \vp }.
\label{tra}
\ee
The left-hand side of \eq{tra} represents the trace of the 
``improved'' energy-momentum tensor
while the right-hand side represents the classical equation of motion for $\vp$.
Thus  the tracelessness of  the ``improved'' energy-momentum tensor 
in the two-dimensional theory invariant under diffeomorphisms is guaranteed 
by the classical equation of motion for $\vp$.

\subsection{Quadratic action}

Under the infinitesimal conformal transformation generated by \rf{hatdel} we have
\bea
\lefteqn{\!\!\!
\LA \hat\delta_\xi  T_{zz}^{(\vp,2)} (\om) \RA=\frac1{2b^2}  \int \d^2 z  
\big\langle q^2 \xi'''(z)+
 \xi'(z) \p^2 \vp(z)\vp(\om) +\xi(z) \p^3 \vp(z)\vp(\om) \big\rangle  \delta^{(2)}(z-\om)}
\non &&= \frac {\xi'''(\om)}{2} \left(\frac{q^2}{b ^2}+ H_{2,1}-H_{3,1} \right)=
 \frac {\xi'''(\om)}{2} \left(\frac{q^2}{b ^2} +\frac13-\frac 16 \right)= 
{\xi'''(\om)}\left( \frac{q^2}{2b ^2}+\frac 1{12}\right) ,~~~~~\non &&
 \label{vp2}
\eea
where we have used \eq{C7}.
Here $1/12$ corresponds to the usual quantum addition $1$ to the central charge.
The right-hand side of \eq{vp2} reproduces the left-hand side of the DDK formula~\rf{DDK1}. 

Analogously for the massive conformal fields we obtain
\bea
\LA \hat\delta_\xi  T_{zz}^{(Y,2)} (\om) \RA&=&\frac1{2b^2}  \int \d^2 z  \big\langle \xi'(z) \p^2 Y(z)Y(\om) +\xi(z) \p^3 Y(z) Y(\om) \big\rangle  \delta^{(2)}_M(z-\om)
\non &=& \frac {\xi'''(\om)}{2} \left( J_{2,1}-J_{3,1} \right)=
\frac {\xi'''(\om)}{2} \left(\frac23-\frac 12 \right)=\frac {\xi'''(\om)}{12}
\label{Y2}
\eea
where we have used \eq{A13}.
The plus sign is for normal statistics. For the regulators with the wrong
statistics the sign changes for minus. This 
explicitly shows how the regulators compensate the quantum part of the central 
charge of $\vp$ so the total one equals the classical value and illustrates the statement that 
we can account for the effect of the regulators by $q\neq 1$ which emerges after
path integration over the regulators. 

Notice that 
the propagators in Eqs.~\rf{vp2} and \rf{Y2} are {\it exact}\/ accounting for the interaction
between $\vp$ and the regulators. This is why renormalized $b^2$ cancels.

\subsection{Quartic action\label{ss:3.3}}

Let us now add  to the quadratic action the $\eps R^2$ term which changes the propagator as
\be
G_\eps(k)=\frac{1}{k^2+\eps k^4},\qquad \delta^{(2)}_\eps(k)=\frac{1}{1+\eps k^2}
\label{pro1}
\ee
and introduces a nontrivial self-interaction of $\vp$.
The computation of $\delta T_{zz}^{(\vp,4)}$ is a bit lengthy 
but easily doable with Mathematica. Equation~\rf{vp2} remains unchanged while 
with the one-loop accuracy we find
\bea
\lefteqn{\eps\LA \hat\delta_\xi T_{zz}^{(\vp,4)} (\om)\RA
=\frac1{b_0^2} \int \d^2 z
 \LA 
2\eps \xi'''(z) [\p\bp\vp(z)-\p\bp\vp(\om)]\vp(\om)
 \right. }\non && 
 ~~~ \left. 
 +\big[ 2\eps \xi''(z) \p^2\bp \vp(z) -6\eps \xi'(z) \p^3\bp \vp(z)-4\eps \xi(z) \p^4\bp \vp(z)\big]\vp(\om) \RA
\delta_\eps^{(2)}(z-\om) \non  && =\frac  {\xi'''(\om) }4 \left( -2 J_{0,1} +2J_{1,1} +6 J_{2,1}
 -4 J_{3,1}\right)
=\frac {\xi'''(\om) }4 \left( -2 \cdot2 +2\cdot 1+6 \frac 23 -4 \frac 12\right) =0, \non &&
\label{vp4}
\eea
where we have used \eq{Jnm} and dropped the logarithmic divergence as prescribed by the
normal ordering. 
Thus the central charge of $\vp$ coincides at one loop 
with that for the quadratic action.

For the regulators we obtain similarly 
\be
\LA \hat \delta_\xi T_{zz}^{(Y,2)}(\om)\RA
=\frac1{b_0^2}\int \d^2 z
 \LA \half \xi'(z)\p^2  Y(z) Y(\om) +\half \xi(z)\p^3  Y(z) Y(\om) \RA
\delta_{\eps,M}^{(2)}(z-\om) =0
\ee
in view of \eq{69} and
\bea
\lefteqn{
\eps\LA \hat \delta_\xi T_{zz}^{(Y,4)}(\om)\RA} \non
&&=\frac1{b_0^2}\int \d^2 z
 \LA \big[ -2\eps \xi''(z)\p^2 \bp Y(z) 
 -6 \eps \xi'(z)\p^3 \bp Y(z) -4 \eps \xi(z)\p^4 \bp Y(z)\big]
 Y(\om)\RA
\delta_{\eps,M}^{(2)}(z-\om) ~~
\non &&=\frac 14 \left(-2 P_{1,1} +6P_{2,1}-4P_{3,1} \right) \xi'''(\om)=
\frac 14\left(-2 \cdot 1 +6\frac 56-4\frac 23 \right) \xi'''(\om) =\frac1{12} \xi'''(\om)
\label{57r}
\eea
at one loop in view of \eq{70}. We see now the same compensation of the 
quantum part in the central charge
of $\vp$ by the regulators as for the quadratic action although in a slightly different way.

We can also perform the computation for the part of $T_{zz}$ in \eq{Tzz} which involves $G$.
For the polynomial in derivatives terms we obtain for the regularization~\rf{pro1}
\bea
\lefteqn{
\LA \hat \delta_\xi \frac{1}{b_0^2}G\eps \left[ (\p\vp)^2 \p\bp\vp-
\p\vp \p(\e^{-\vp}\p \vp \bp \vp)+ \p^2(\e^{-\vp}\p \vp \bp \vp)-\p(\p\vp \e^{-\vp} \p\bp \vp)\right]\RA}\non&&= 
\frac{1}{b_0^2}G\eps \int \d^2 z
\LA \xi'''(z)\big[-\p\bp\vp(z)\vp(\om)   
-3\p\vp(\om)\bp\vp(\om)\big]  
-2 \xi''(z) \p^2\bp \vp(z) \vp(\om)
\RA \delta_\eps^{(2)}(z-\om) \non &&=
-\frac 32 G \xi'''(\om) \int \d k^2 \frac{\eps}{1+\eps k^2} .
\label{tttp}
\eea
For the nonlocal term in \rf{Tzz} we analogously find 
\bea
\LA \hat \delta_\xi \left(-\frac{1}{b_0^2}G\eps \frac 1\bp \p^2 (\bp\vp \e^{-\vp} \p\bp \vp)\right)\RA 
&= &
-\frac{1}{b_0^2}G\eps \int \d^2 z\, \xi'''(z)
 \LA\p\bp\vp(z)\vp(\om) + \p\bp\vp(\om)\vp(\om)\RA  \non &&\hspace*{-3cm}\times
\delta_\eps^{(2)}(z-\om) = \frac 12 G \xi'''(\om)
+\frac 12 G \xi'''(\om) \int \d k^2 \frac{\eps}{1+\eps k^2} .
\label{tttnlp}
\eea
In contrast to the average of the nonlocal (last) term in the classical formula \rf{delTza} 
now a nonvanishing finite contribution arises.

The sum of \rf{vp2}, \rf{vp4}, \rf{tttp} and \rf{tttnlp} precisely reproduces the central charge \rf{cphi} at one loop.

For a future use I present also the exact formula for the conformal variation of
the normal-ordered $ T_{zz}^{(\vp,4)}$
\bea
\lefteqn{\LA\hat\delta_\xi T_{zz}^{(\vp,4)} (\om)\RA 
= 2 \frac1{b^2} \int \d^2 z 
\Big\langle \!\e^{-\vp(\om)} 
\big\{- \xi'''(z) \p\bp\vp (z) +\xi''(z) \p\vp(z)\p\bp \vp(\om)} \non &
+&\!\xi'(z)\big[2\p^3\bp \vp(z)-\p^2\vp(z)\p\bp\vp(\om)-
\p\bp \vp(z)(\p\vp(\om))^2-\p\vp(z)\p\vp(\om)\p\bp\vp(\om)\big]\non &
+&\!\xi(z)\big[\p^4\bp \vp(z)-\p^3\vp(z)\p\bp\vp(\om)-
\p^2\bp \vp(z)(\p\vp(\om))^2\!+\p\vp(z)\p^2\vp(\om)\p\bp\vp(\om)\big] \big\}\!
\Big\rangle \delta_\eps^{(2)}(z-\om).\!\non&&
\label{ppp}
\eea
Equation~\rf{vp4} is its expansion to quadratic order in $\vp$.
Equation~\rf{ppp} can be possibly useful to show, manipulating with the derivatives,
 that its right hand side in fact vanishing like
\rf{vp4}. That would prove the universality to all loops.

\subsection{Six-derivative action}

Under the infinitesimal conformal transformation we have for the six-derivative action at one loop
the following variation of~\rf{T6}:
\bea
\LA \hat \delta_\xi T_{zz}^{(\vp,6)}(\om) \RA &=&\frac {1}{b_0^2}
\int \d^2 z \LA\Big[-8 \xi^{(4)}(z) \p\bp^2\vp(z)
-8\xi'''(z)\p^2\bp^2\vp(z)+16\xi''(z) \p^3\bp^2\vp(z)\right.
\non
&&\hspace*{1.cm}
\left.+48\xi'(z)\p^4\bp^2 \vp(z) +24 \xi(z) \p^5 \bp^2 \vp(z)
\Big] \vp(\om) \RA \delta_{\eps}^{(2)}(z-\om).
\label{vp6}
\eea

If we consider the six-derivative action
\be
S[\vp]=S^{(\vp,2)} + 2\eps S^{(\vp,4)}+\eps^2  S^{(\vp,6)}
\label{S6d}
\ee
the propagator will be of the form \rf{Geps} with $m=2$.
Substituting in \rf{vp2}, \rf{vp4}, \rf{vp6} we find
\bea
\LA \hat \delta_\xi T_{zz}^{(\vp,2)}(\om) \RA &=&
\frac {\xi'''(\om)}{2} \left(\frac{q^2}{b ^2}+ H_{2,2}-H_{3,2} \right) =
\frac {\xi'''(\om)}{2} \left(\frac{q^2}{b ^2}+ \frac3{10}-\frac2{15} \right) 
\non &=&
\xi'''(\om) \left(\frac{q^2}{2b ^2}+ \frac1{12} \right), 
\eea
\bea
2 \eps\LA \hat \delta_\xi T_{zz}^{(\vp,4)}(\om) \RA &=&
2\frac  {\xi'''(\om) }4 \left( -2 J_{0,2} +2J_{1,2} +6 J_{2,2} -4 J_{3,2}\right) \nonumber\\
&=&2\frac  {\xi'''(\om) }4 \left( -2  \frac23+2\frac13 +6\frac 15 -4 \frac{2}{15}\right)=0,
\eea
\bea
\eps^2\LA \hat \delta_\xi T_{zz}^{(\vp,6)}(\om) \RA &=&
\xi'''(\om)\left(-\frac 12 Q_{0,2}-Q_{1,2}+3 Q_{2,2}-\frac32 Q_{3,2}\right) \non
&=&\xi'''(\om)\left(-\frac 12\cdot \frac13-\frac13+3 \frac3{10}-\frac32 \cdot\frac4{15}\right) 
=0.
\eea
We have thus obtained the same result as for the quadratic and quartic actions.

One more argument in favor of the universality, \ie independence of the central charge of $\vp$
on the form of the action, are the identities
\begin{subequations}
\bea
H_{2,m}-H_{3,m} &=&\frac 16, \\
 -2 J_{0,m} +2J_{1,m} +6 J_{2,m} -4 J_{3,m}&=&0,\\ 
 - Q_{0,m}-Q_{1,m}+3 Q_{2,m}-\frac32 Q_{3,m}&=&0
\eea
 \label{ide}
\end{subequations}
\!satisfied for the propagator \rf{Geps} associated with the regularization by yet higher
derivatives.

The formula analogous to \rf{vp6} can be derived  also for the regulators 
\bea
\LA \hat \delta_\xi T_{zz}^{(Y,6)}(\om) \RA &=&\frac 1{b_0^2}
\int \d^2 z \LA\Big[ 
8\xi'''(z)\big(\p^2\bp^2 Y(z)-\p^2\bp^2 Y(\om) \big)+32\xi''(z) \p^3\bp^2 Y(z) \right.
\non
&&\hspace*{.8cm}\left.+48\xi'(z)\p^4\bp^2 Y(z) +24 \xi(z) \p^5 \bp^2 Y(z)
\Big] Y(\om)  \RA\delta_{\eps,M}^{(2)}(z-\om).~~~~
\eea
The massive analogue of \eq{Qnm} involves the terms $\propto M/\sqrt{\eps}$ which
make analisis more complicated.

\section{Universality of the six-derivative action\label{s:4}}

\subsection{$R\Delta R$}

The most general six-derivative action of the form~\rf{Sgen} reads
\be
S[\vp]=S^{(\vp,2)}+(\eps+a^2)S^{(\vp,4)}+\eps a^2 S^{(\vp,6)}
\label{S6dg} 
\ee
involving the term $R\Delta R$.
The action~\rf{S6d} corresponds to $a^2=\eps$.
The six-derivative action~\rf{S6dg} is associated to the regularization 
\be
G_{\eps,a}(k)=\frac{1}{k^2(1+\eps k^2)(1+a^2 k^2)^{m-1}},\qquad 
\delta^{(2)}_{\eps,a}(k)=\frac{1}{(1+\eps k^2)(1+a^2 k^2)^{m-1} }
\label{Gepsa}
\ee
with $m=2$.

For an arbitrary ratio $a^2/\eps$ we write
\begin{subequations}
\bea
\frac1{b_0^2}
\int \d^2 z f(z) \LA \p^{n} \vp(z)  \vp(\om) \RA
\delta_{\eps,a}^{(2)}(z-\om)&= &(-1)^n H_{n.m} \big(\frac {a^2}{\eps} \big)\p^n f(\om), 
\label{epsa0f} \\
\frac1{b_0^2}
\int \d^2 z f(z) \LA -4 \eps \p^{n+1} \bp \vp(z)  \vp(\om) \RA
\delta_{\eps,a}^{(2)}(z-\om)&= &(-1)^n J_{n,m} \big(\frac {a^2}{\eps} \big)
\p^n f(\om), 
\label{epsa1f} \\
\frac1{b_0^2}
\int \d^2 z f(z) \LA 16 \eps a^2 \p^{n+2} \bp^2 \vp(z)  \vp(\om) \RA
\delta_{\eps,a}^{(2)}(z-\om)&=& (-1)^n Q_{n,m}  \big(\frac {a^2}{\eps} \big)
\p^n f(\om),
\label{epsa2f}
\eea
\label{epsaf}
\end{subequations}
\!where the functions on the right-hand side are like
\be
J_{1,2} \big(\frac {a^2}{\eps} \big) =
\frac{\eps\left(a^4-\eps^2-2a^2\eps \log\big(\frac {a^2}{\eps} \big)\right)}{(a^2-\eps)^3}.
\ee
They obey the identities
\begin{subequations}
\bea
H_{2,m}\big(\frac {a^2}{\eps} \big)-H_{3,m}\big(\frac {a^2}{\eps} \big) &=&\frac 16, \\
 -2 J_{0,m}\big(\frac {a^2}{\eps} \big) +2J_{1,m}\big(\frac {a^2}{\eps} \big)
  +6 J_{2,m}\big(\frac {a^2}{\eps} \big) -4 J_{3,m}\big(\frac {a^2}{\eps} \big)&=&0,\\ 
 - Q_{0,m}\big(\frac {a^2}{\eps} \big)-Q_{1,m}\big(\frac {a^2}{\eps} \big)
 +3 Q_{2,m}\big(\frac {a^2}{\eps} \big)-\frac32 Q_{3,m}\big(\frac {a^2}{\eps} \big)&=&0,
\eea
\end{subequations}
\!generalizing \rf{ide} to $a^2\neq \eps$ and
illustrating the universality at one loop.

\subsection{$R^3$}

The consideration of the contribution to the central charge of $\vp$
from the $R^3$ term in the emergent action, 
 which is not of the type shown in \eq{Sgen}, is pretty much similar.
 The difference between the $R^3$ term and the $R\Delta R$ term considered in
 the previous subsection is that its expansion in $\vp$ starts from $\vp^3$
 rather than $\vp^2$. Then only the piece of the contribution of $R^3 $ to
 the ``improved'' energy momentum tensor of the form
 \begin{equation}
\widetilde T^{(\varphi)}_{zz}= \frac{12}{b_0^2}\partial^2
(\partial\bar\partial\varphi)^2+{\cal O}(\varphi^3)
 \label{delT3}
 \end{equation}
which is linked to the shift~\rf{Rshift}
may give a nonvanishing result at one loop. The piece like $\vp^3$  
which yields $\vp^2$ after acting by the variational derivative $\delta/\delta \vp$
from the term with $\xi'$ in \rf{hatdel} vanishes after the averaging because the
variation of a normal product is again a normal product (see the footnote$^{\ref{foo:2}}$).
 
The variation of \rf{delT3} under the infinitesimal conformal transformation~\rf{hatdel} is
easily calculable
\begin{eqnarray}
\varepsilon^2\left\langle \hat \delta_\xi\widetilde T_{zz}^{(\vp)}(\omega) \right\rangle
&= &\frac {12}{b_0^2}\int {\rm d}^2 z \left\langle\Big[ 2\varepsilon^2\xi'''(z)
\partial^2\bar\partial^2\vp(z)+2\varepsilon^2\xi''(z) \partial^3\bar\partial^2\varphi(z)
\Big] \vp(\omega)\right\rangle \delta_{\varepsilon,a}^{(2)}(z-\omega) \nonumber \\* &=&
\frac32\xi'''(\omega) \left[ Q_{0,m}\big(\frac {a^2}{\varepsilon} \big)
+ Q_{1,m}\big(\frac {a^2}{\varepsilon} \big) \right]=0.
\label{vpR3}
\end{eqnarray}
It vanishes thus supporting the universality.

From the above analysis of the contribution of the term $R^3$ to the central charge of $\vp$
at one loop
it becomes clear that most of yet higher-derivative terms in the effective action do not
contribute for the trivial reason -- too many $\vp$'s. In addition to the action~\rf{Sgen} only the terms 
$\eps^{n+2} R\Delta^k R \Delta ^{n-k} R$  with $n\geq 1$, $0\leq k \leq n$ are to be
analyzed like the $R^3$ term above.
 
\section{Discussion\label{s:5}}

The above computations confirm the result~\rf{cphi} for the one-loop  central charge
of $\vp$ for the action~\rf{inva}.
They are also useful for studying the universality of the higher-derivative actions.
We have explicitly shown at one loop the universality of the central charge for the quartic and
six-derivative actions.

\begin{figure}
\centerline
{\includegraphics[width=7.8cm]{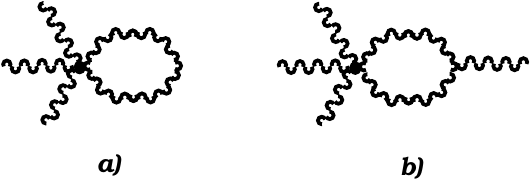}}
\caption{One-loop renormalization of $\e^{\vp}$ whose position is depicted by the dot. 
The wavy lines represent $\vp$.}
\label{fi:fig3}
\end{figure}

How the universality works for the action~\rf{Sgen}
is easily seen  in the one-loop computation of
the renormalization of $\e^\vp$ given by the diagrams in Fig~\ref{fi:fig3}.
The diagram in Fig~\ref{fi:fig3}$a$ is  obviously universal with logarithmic accuracy.
The diagram  in Fig~\ref{fi:fig3}$b$ involves the propagator $k^2 F(\eps k^2)$ and
the triple vertex which comes from the variation
of \rf{Sgen} with respect to $\vp$ resulting in $\eps k^4F'(\eps k^2)$. Its contribution~\cite{Mak21}
\be
{\rm Fig.~\ref{fi:fig3}b}=
\frac {\e^\vp}2
\times 8\pi b^2_0 \vp \int \frac {\d^2 k}{(2\pi)^2} 
\frac {\eps k^4 F'(\eps k^2)}{[k^2F(\eps k^2)]^2}=  
\frac 1{F(0)}\e^\vp b^2_0 \vp =  \e^\vp b^2_0 \vp 
\label{vvvgen}
\ee
does not depend on the choice of the function $F$.

In fact my motivation to analyze the singular products was to go 
beyond the one-loop approximation for the central charge of $\vp$. This may be doable by the
described method
of singular products if to prove the vanishing of $\LA \hat \delta_\xi T_{zz}^{(\vp,2N)}\RA $
to all loops like it was done for $N=2,3$ at one loop. Then we may expect the 
exact central charge to be
\be
c^{(\vp)}= \frac{6q^2}{b^2}+1+6 G q.
\label{cphie}
\ee
I hope to return to this issue elsewhere.

\subsection*{Acknowledgement}

This work was supported by the Russian Science Foundation (Grant No.20-12-00195).



\appendix

\section{Derivation of emergent action for the Nambu-Goto string\label{AppA}}

The path integration over $X^\mu$ for the Nambu-Goto action~\rf{SNG} can be performed 
using  the DeWitt-Seeley expansion of
the operator $(\sqrt{g})^{-1}\p_a \lambda^{ab}\p_b $. Splitting the operator into the 
gravitational and ``electromagnetic'' parts, we have
\be
{\cal O}\equiv\frac 1 {\sqrt{g}}\p_a \lambda^{ab} \p_b = h^{ab} \p_a\p_b + A^a \p_a,\qquad
h^{ab}=\frac{\e^{-\vp}}{\sqrt{\hg}} \lambda^{ab}
,\quad A^b=\frac {\e^{-\vp} }{\sqrt{\hg}} \p_a \lambda^{ab}.
\label{defg}
\ee
Using the results known from~\cite{deWitt,Gil75} (see \cite{Vas03} for a review), we write the expansion
\be
\LA \Big| \e^{\tau {\cal O} }
\Big|  \RA = \frac 1{4\pi\tau}
 +\frac1{4\pi} \left(\frac 16{R} +E
 \right)
 +\frac{\tau}{120\pi}
\left( \Delta R+\frac 12 R^2
\right)+
{\cal O}(\tau^2),
\label{2.5}
\ee
where $R$ and $\Delta$ are the scalar curvature and the Laplacian for the metric tensor
$h_{ab} $ defined in \eq{defg}. We have dropped here the term $\propto\tau A^2$
because $A\sim\tau$ with $\tau$ being the proper-time UV cutoff, as it will be momentarily seen.

To find the emergent action for the fluctuating fields $\vp$ and 
$\delta \lambda^{ab}=\lambda^{ab}-\sqrt{\hg} \hg^{ab}$, we use
\be
h^{ab}= \hg^{ab}+ \delta h^{ab}, \qquad 
\delta h^{ab}=-\hg^{ab} \vp+\frac{\delta \lambda^{ab}}{\sqrt{\hg}}
\ee
or
\be
 \delta h^{z\bz}=2 \delta \lambda^{z\bz}-2\vp,\quad
  \delta h^{zz}=2 \delta \lambda^{zz},\quad \delta h^{\bz\bz}=2 \delta \lambda^{\bz\bz}
 \ee
in the complex coordinates $z$ and $\bz$.
To quadratc order in the fluctuations,
we can use the formulas in an inertial frame
\bea
R&=&-\p_a \p_b h^{ab}+ h_{ab} h^{cd} \p_c \p_d h^{ab} =  
 -\p^2 \delta h^{zz} -\bp^2 \delta h^{\bz\bz}+2 \p \bp \delta h^{z\bz}, \\
 E&=&-\frac 12 \left(\p_a A^a -\p_a\p_b h^{ab}+\frac 12 \hg_{ab}\hat\Delta h^{ab}
 \right)
\eea
which yields 
\be
\frac 16 R +E = 
-\frac 13\left(\p^2 \delta \lambda^{zz} +\bp^2 \delta \lambda^{\bz\bz}
+4 \p \bp \delta \lambda^{z\bz} +2 \p\bp \vp \right).
\ee

Using the fact that the variational derivative of the emergent action with respect to $\vp$ is equal to~\rf{2.5}
and covariantizing, we find the following contribution from  $X^\mu$  to the emergent action:
\bea
{\cal S}_X[\vp, \lambda^{ab}]
&=&\frac d2 \int \left[- \frac {\e^\vp}{4\pi\tau \sqrt{\det{\lambda^{ab}}}}
+\frac1{12\pi}\left( \vp\p\bp\vp +4
\lambda^{z\bz} \p\bp \vp+
\lambda^{zz}\nabla\p \vp  +\lambda^{\bz\bz}\bar\nabla\bp\vp \right)
\right.\non &&\hspace*{1.2cm}\left.
+\frac{\tau}{15\pi} \e^{-\vp} (\p\bp \vp)^2 \right] +{\cal O}(\tau^2).
\label{SXa}
\eea
Here $\nabla = \p-\p \vp$ is the covariant derivative in the conformal gauge.
Equations~\rf{SXb} and \rf{SXa} perfectly agree.%
\footnote{The last term here is 2/3 of the one for the Pauli-Villars regularization as is prescribed
by Apppendix~A of \cite{Mak23b}.\label{ffot}}

To describe fluctuations, we expand about $\lambda^{ab}=\sqrt{\hg} \hg^{ab}$  when
\be
\lambda ^{z\bz}=1+\delta \lambda ^{z\bz},\qquad
\lambda ^{zz}=\delta \lambda ^{zz},\qquad
\lambda ^{\bz\bz}=\delta \lambda ^{\bz\bz}
\ee
to get 
\be
 \frac {1}{ \sqrt{\det{\lambda^{ab}}}}=1-\delta \lambda ^{z\bz}
 + (\delta \lambda ^{z\bz})^2 +\frac12 \delta \lambda ^{zz}\delta\lambda ^{\bz\bz} +
 {\cal O}(\delta \lambda)^3 .
\label{detla}
\ee
The integrals over $\delta \lambda^{zz}$ and 
$\delta \lambda^{\bz\bz}$ have then saddle points at
\be
\delta \lambda^{zz} =\frac {2}{3} \tau \e^{-\vp}\bar\nabla \bp \vp,\qquad 
\delta \lambda^{\bz\bz} =\frac {2}{3} \tau \e^{-\vp} \nabla \p \vp
\label{sp1}
\ee
which demonstrates that $\delta \lambda \sim \tau$, justifying the expansion in $\delta \lambda$
and the saddle point.

It is slightly different with $\delta \lambda^{z\bz}$ because of the linear term in \eq{detla}.
It simply renormalizes the bare string tension in  the classical part of the Nambu-Goto action
(the last term on the right-hand side of \eq{SNG}). 
We thus have at the saddle point
\be
\delta \lambda^{z\bz}= \tau \left(\frac 23  \e^{-\vp}\p\bp\vp -\frac{1}{2d\alpha'_R}
\right),
\label{sp2}
\ee
where the second term in the brackets can be omitted for finite $\alpha'_R$ as $\tau\to0$.

Inserting the saddle-point values~\rf{sp1}, \rf{sp2} into the 
action~\rf{SXa}, we find
\be
{\cal S}_X[\vp]= \frac d{2}\int \left\{- \frac {\e^\vp}{4\pi\tau }
+\frac1{12\pi} \vp\p\bp\vp
+
{\tau}\e^{-\vp}\left[\frac8{45\pi}(\p\bp\vp)^2
+\frac1{18\pi} \nabla \p \vp \bar\nabla \bp \vp \right]
\right\}+{\cal O}(\tau^2).
\ee
The final step to obtain the quartic in derivatives part of the action~\rf{inva} is to use
 the identity
\be
\e^{-\vp} \nabla\p  \vp \bar\nabla\bp \vp 
= \e^{-\vp}\left[
(\partial\bar\partial \vp)^2 +\partial \vp\bar\partial \vp \partial\bar\partial \vp
\right]  
+\partial \left(\e^{-\vp} \partial \vp  \bar\nabla\bp \vp 
\right)-\bar\partial (\e^{-\vp} \partial \vp \partial\bar \partial \vp),
\label{ide4}
\ee
to get
\be
{\cal S}_X[\vp]= \frac d{2}\int \left\{- \frac {\e^\vp}{4\pi\tau }
+\frac1{12\pi} \vp\p\bp\vp
+
{\tau}\e^{-\vp}\left[\frac7{30\pi}(\p\bp\vp)^2
+\frac1{18\pi}  \p \vp  \bp \vp \p\bp\vp\right]
\right\}+{\cal O}(\tau^2).
\label{A.15}
\ee

Summing \rf{A.15} with the contribution from the ghosts (see the footnote${}^{\ref{ffot}}$)
\be
{\cal S}_{\rm gh}[\vp]= \int \left[ \frac{\e^{\vp}}{4\pi\tau}
 -\frac {13}{12\pi}\vp \p\bp\vp -\frac{11\tau}{15\pi }\e^{-\vp}  (\p\bp \vp)^2 \right]
 + {\cal O}(\tau^2),
\label{SghA}
\ee
we obtain \eq{inva} previously derived~\cite{Mak21} for the Pauli-Villars regularization.  

\section{List of formulas for the singular products\label{AppB}}

The simplest singular product
\be
\int \d^2 z\, \xi(z) \LA \partial^n \vp(z) \vp(0) \RA 
\delta^{(2)}(z) =(-1)^{n}  \frac{2}{n(n+1)} \xi^{(n)}(0) 
\label{C7}
\ee
emerges in a free CFT, where the propagator is
\be
\LA \vp(z) \vp(0) \RA =8\pi G_0(z), \qquad
G_0(z) =-\frac1{2\pi} \log\Big(\sqrt{z\bz} \mu\Big)
\label{G0}
\ee
and $\mu$ represents an infrared cutoff. Equation \rf{C7} can be derived by the formulas
\be
\delta^{(2)}(z) =\bp \frac {1}{\pi z},\qquad
\frac 1{z^n} \bp \frac 1z=(-1)^n \frac1{(n+1)!} \p^n \bp \frac 1z.
\label{B2}
\ee

Equation~\rf{C7} can be alternatively derived introducing 
 the regularization by $\eps$ via higher derivatives.  In momentum space we define
\be
G_{\eps}(k)=\frac{1}{k^2(1+\eps k^2)^m},\qquad 
\delta^{(2)}_\eps(k)=\frac{1}{(1+\eps k^2)^m }.
\label{Geps}
\ee
We then have
\bea
\frac1{b_0^2}
\int \d^2 z f(z) \LA \p^{n} \vp(z)  \vp(\om) \RA
\delta^{(2)}_\eps(z-\om)= (-1)^n H_{n.m}
\p^n f(\om), \non
 H_{n.m}=\frac{2^{2m}\Gamma(m+1/2) \Gamma(n+m)} {\sqrt{\pi}\, n\,\Gamma(n+2m)}
 \hspace*{3cm}
\eea
and
\bea
\frac1{b_0^2}
\int \d^2 z f(z) \LA -4 \eps\p^{n+1} \bp \vp(z)  \vp(\om) \RA
\delta^{(2)}_\eps(z-\om)= (-1)^n J_{n.m}
\p^n f(\om), \non
 J_{n,m}=\frac n{2m-1} H_{n,m}=\frac{2^{2m-1}\Gamma(m-1/2)
\Gamma(n+m)} {\sqrt{\pi} \Gamma(n+2m)}.\hspace*{1cm}
\label{Jnm}
\eea
For $m=1$ this gives
\be
H_{n,1}=\frac 2{n(n+1)}, \qquad J_{n,1}=\frac2{(n+1)},
\label{m=1}
\ee
reproducing \eq{C7}.

For the six-derivative action we shall also need
\be
\frac1{b_0^2}
\int \d^2 z f(z) \LA 16 \eps^2\p^{n+2} \bp^2 \vp(z)  \vp(\om) \RA
\delta^{(2)}_\eps(z-\om)= (-1)^n Q_{n.m}
\p^n f(\om)
\ee
with
\bea
 Q_{0,m}=Q_{1,m}=\frac1{(m-1)(2m-1)},&\qquad&
 Q_{2,m}=\frac{3(m+1)}{2(m-1)(4m^2-1)},\non
  Q_{3,m}=\frac{(m+2)}{(m-1)(4m^2-1)},&\qquad & 
  Q_{n,2}=\frac{2(n+1)}{(n^2+5n+6)}.
\label{Qnm}
\eea

 
 In the case of the  massive conformal fields (regulators) we have
 \be
\int \d^2 z\, \xi(z) \LA \p^n Y(z) Y(0) \RA \delta^{(2)}_M(z)=
(-1)^{n} J_{n,1}\xi^{(n)}(0)
\label{A13}
\ee
for the free massive propagator
\be
\LA  Y(z) Y(0) \RA =8\pi G_M(z)=4 K_0\left( M \sqrt {z\bz} \right)
,\qquad G_M(k)=\frac 1{k^2+M^2}
\ee
and an ad hoc regularization of the delta function by 
the large-$M$ limit of
\be
\delta^{(2)}_M(z)= \frac{M^2}{2\pi}K_0\big(M\sqrt{z \bz}\big),\qquad
\delta^{(2)}_M(k)=\frac{M^2}{k^2+M^2}
\ee
what is natural if $Y$ is  the Pauli-Villars regulator. 
The same $J_{n,1}$ as shown in \eq{m=1} appears in \eq{A13}.
 
For the quartic derivative we define
 \be
\LA Y(-k) Y(k)\RA= 8\pi b_0^2  G_{\eps,M},\quad
 G_{\eps,M}= \frac{(k^2+M^2)^{m-1}}{(k^2+M^2+\eps k^4)^m},\qquad 
 \del_{\eps,M}^{(2)}=\frac{(k^2+M^2)^m}{(k^2+M^2+\eps k^4)^m}
 \ee
to reproduce \rf{Geps} as $M\to0$. In the opposite limit $M\to\infty$ we find
 \be
 8 \pi \int \d^2 z f(z) \p^{n} G_{\eps,M}(z-\om)  \del_{\eps,M}^{(2)}(z-\om)
 \stackrel{M\to \infty} \to 0,
 \label{69}
 \ee
 \bea
 8 \pi \int \d^2 z f(z) (-4 \eps\p^{n+1}\bp) 
 G_{\eps,M}(z-\om)  \del_{\eps,M}^{(2)}(z-\om)
 \stackrel{M\to\infty} \to 
 (-1)^n P_{n,m} \p^n f(\om), 
 \non 
 P_{0,m}= P_{1,m}=\frac1{(2m-1)},\quad   P_{2,m}=\frac{(2m+3)}{2(4m^2-1)},\quad
 P_{3,m}=\frac{2}{(4m^2-1)} 
 .
  ~~~~~~~~
 \label{70}
 \eea

\vspace*{3mm}
\paragraph{Note added in the proof:} Equation~\rf{cphie} has been derived recently 
in \cite{Mak23e}.

\end{document}